\documentclass[%
 reprint,
 amsmath,amssymb,
 aps,
]{revtex4-2}
\usepackage{upgreek}
\usepackage{multirow}
\usepackage{graphicx}% Include figure files
\usepackage{dcolumn}% Align table columns on decimal point
\usepackage{bm}% bold math
\usepackage{soul}
\usepackage{xcolor}
\newcommand*\rv[1]{#1}
\newcommand*\HL[1]{#1}
\newcommand*\rrv[1]{#1}
\newcommand*\rrrv[1]{#1}

\begin{document}

\preprint{APS/123-QED}

\title{Spatial Filtering with Nonlocal Non-Hermitian Metasurfaces} 

\author{Biao Chen}
\email{biao.chen@aalto.fi}
\author{Mikael Reichler}
\author{Radoslaw Kolkowski}
 \email{radoslaw.kolkowski@aalto.fi}
\author{Andriy Shevchenko}
 \email{andriy.shevchenko@aalto.fi}
\affiliation{%
 Department of Applied Physics, Aalto University, P.O.Box 13500, Aalto FI-00076, Finland
}%

\date{\today}% It is always \today, today,

\begin{abstract}
Spatial filtering of optical fields has widespread applications ranging from beam shaping to optical information processing. However, conventional spatial filters are bulky and alignment-sensitive. Here, we present nonlocal non-Hermitian metasurfaces that can act as exceptionally effective optical spatial filters while being highly compact and insensitive to both lateral and longitudinal displacements. The metasurface design is based on a resonant waveguide grating in which radiative losses of the modes are tailored to realize a symmetry-protected bound state in the continuum in the middle of a non-Hermitian flat band. Using this design, we propose a compact spatial filtering device operating over an angular range of approximately 1 degree around normal incidence. In addition to being ultrathin and robust against translational misalignment, the proposed metasurfaces are easy to manufacture, which makes them an attractive alternative to conventional spatial filters, holding a potential to become a widely used optical component.
\end{abstract}

\maketitle

%\tableofcontents

\section{Introduction}
Optical metasurfaces are two-dimensional arrays of nanostructures that can be designed to provide almost arbitrary control over light fields~\cite{schulz2024roadmap}. Recently, the concept of nonlocal metasurfaces has attracted significant attention, unlocking alternative ways of tailoring light through its coupling to spatially extended collective modes of metasurfaces~\cite{kolkowski2023nonlinear,overvig2022diffractive,shastri2023nonlocal}, e.g., making use of guided-mode resonances (GMRs) \cite{wang1993theory, quaranta2018recent}. These modes allow one to directly alter the plane-wave components (spatial Fourier spectrum) rather than local characteristics of optical fields~\cite{ouyang2022optical}. Designing such metasurfaces requires appropriate engineering of their photonic band structures, with particular focus on the efficiency of coupling between confined nonlocal modes and free-space radiation. For example, highly symmetric metasurfaces can support modes known as optical bound states in the continuum (BICs)~\cite{hsu2016bound} that are decoupled from free-space radiation completely. Owing to their unlimited quality factors and strong local field enhancement, BICs have recently been considered for diverse light-shaping applications~\cite{azzam2021photonic,kang2023applications}. On the other hand, coupling between modes with high and low radiative loss allows one to implement the concept of parity-time ($\mathcal{PT}$) symmetry and non-Hermitian physics~\cite{zhen2015spawning}. In this approach, the dispersion of photonic modes is modified through the formation of flat bands and exceptional points, which has been considered as a platform for wavefront shaping~\cite{cerjan2016exceptional} and controlling light polarization~\cite{kolkowski2021pseudochirality}. Photonic flat bands 	can also be achieved by tailoring the spatial symmetry of the structures \cite{leykam2018perspective} and they have long been considered for slow-light applications \cite{yang2023photonic}. On the other hand, it has recently been shown that non-Hermitian photonic band features can also emerge in time-modulated systems \cite{park2022revealing,asgari2024theory}. \rrv{However, despite intensive research in the field of non-Hermitian metasurfaces, no practical implementation of optical devices exploiting both nonlocality and non-Hermiticity has been reported so far. In particular, spatial filters based on such metasurfaces have not been introduced.}

In this work, we present a simple design of a nonlocal non-Hermitian metasurface that acts as an exceptionally efficient optical spatial filter. The metasurface consists of a metal grating placed on a dielectric slab waveguide. Its filtering capabilities are enabled by the GMRs forming a peculiar photonic band structure with a $\Gamma$-point BIC in the middle of a non-Hermitian flat band~\cite{deng2022extreme,kolkowski2023bound,niu2024metasurface}. The presence of this BIC is enforced by the two-fold rotational symmetry of the grating~\cite{zhen2014topological,bai2024recovery}, while the non-Hermitian flat band results from frequency degeneracy of two modes differing in their radiative loss. Such a band structure provides a narrow angular range close to normal incidence that can efficiently be used for either low-pass or high-pass spatial filtering, depending on the polarization of incident light and the type of designed GMRs, either transverse electric (TE) or transverse magnetic (TM). \rrv{To explain the band structure of the designed metasurface, we use a well-known two-state model that results in a flat band and exceptional points relevant to the achieved spatial filtering by the system.} While a single metasurface acts as a one-dimensional spatial filter (along one direction in the Fourier space), a combination of two such metasurfaces with a half-wave plate provides the full two-dimensional filtering, which we demonstrate numerically with a multimode beam and a grayscale image. The use of non-Hermitian flat bands distinguishes our metasurfaces from other structures used to
spatially filter optical fields~\cite{silva2014performing, golovastikov2014resonant,maigyte2015spatial,roberts2018optical,zhou2019optical,qian2019all,wesemann2021meta, wan2020optical, komar2021edge, bi2023wideband, cotrufo2023dispersion, cotrufo2023polarization, tanriover2023metasurface}. Furthermore, as opposed to conventional spatial filters composed of two lenses and a pinhole, our filters are compact and insensitive to translational misalignment. We also show that the filtering performance of our systems is only weakly affected by variations of the structure parameters, which makes them feasible for fabrication and practical use. Overall, the unique properties of the proposed approach make it a promising alternative to the traditional methods of spatial filtering.

\begin{figure}[t!]
	\includegraphics[width=\columnwidth]{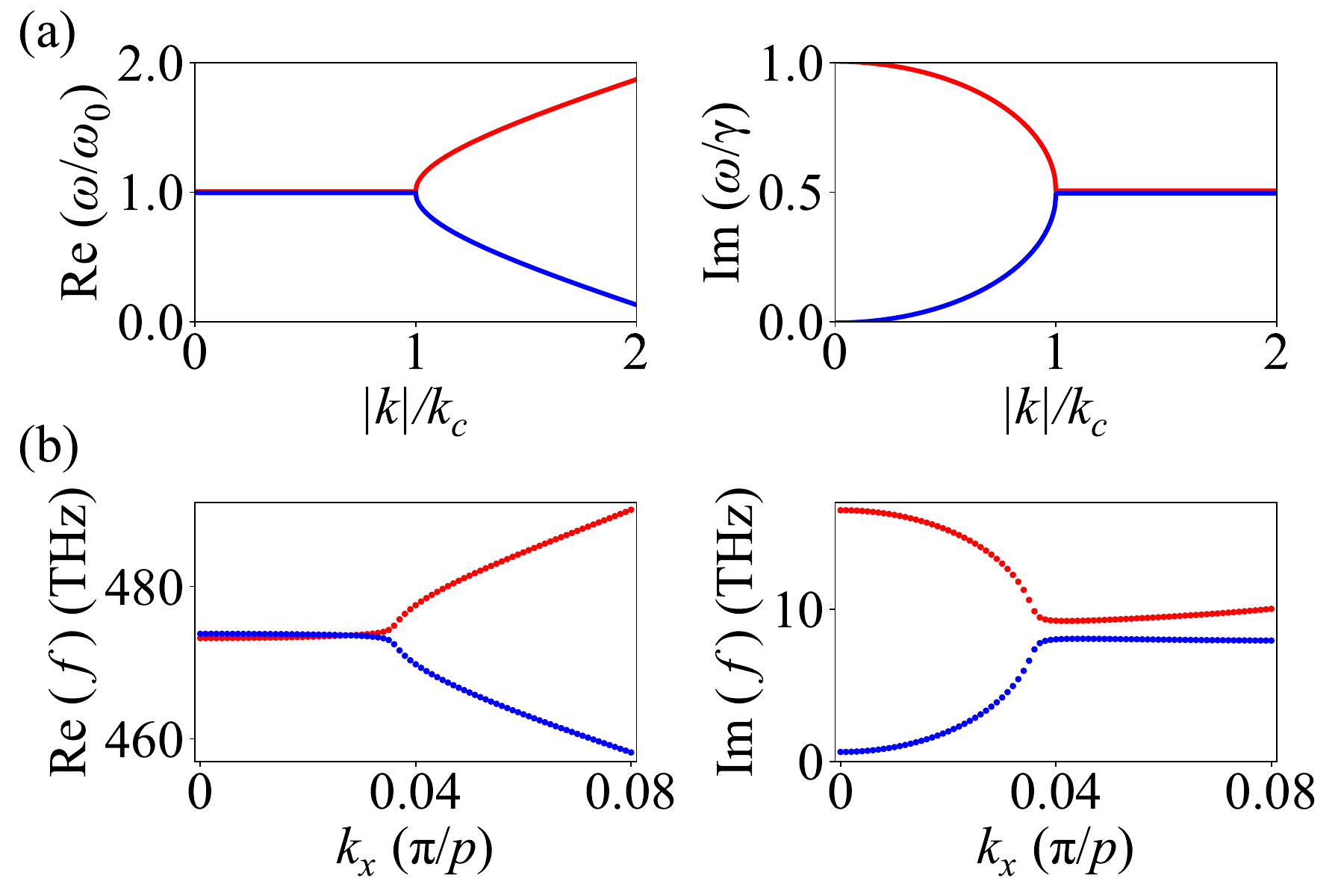}
	\caption{(a) Real and imaginary parts of the eigenfrequencies of a non-Hermitian photonic system as a function of the normalized propagation constant. (b) Numerically calculated band structure of a nonlocal non-Hermitian metasurface shown in Fig. 2 with parameter values shown in Table I. The real and imaginary parts of the eigenfrequencies ($f$) are plotted for TM-polarized GMRs as functions of $k_x$ ($k_y = 0$, see Fig. \ref{Fig2}).} \label{Fig1}	
\end{figure}

\section{Approach}\label{sec2}

The design principle of the proposed metasurfaces can be explained by considering a one-dimensional photonic system in which a lossy mode (i.e., having a significant radiation loss) is coupled to a lossless mode (i.e., a BIC that is symmetry-mismatched with any external propagating plane wave). In the following simple model, we neglect any absorption loss and call the lossy and lossless modes ``bright'' and ``dark'', respectively. In the absence of coupling, the eigenfrequency of the dark mode, $\omega_d$, is purely real, whereas the eigenfrequency of the bright mode can be expressed as $\omega_b+i\gamma$, where $\gamma$ corresponds to the loss rate due to radiation. We assume that the real parts of the eigenfrequencies of both modes are equal, i.e., $\omega_d$ = $\omega_b$ = $\omega_0$. The coupling between these modes near the $\Gamma$-point can be introduced in the form of a momentum-dependent term, $v_gk$, corresponding to a linear dispersion of the modes, where $v_g$ is the mode group velocity in the absence of radiation loss~\cite{zhen2015spawning}. The propagation constant of the modes, $k$, can be considered to match the transverse wavevector of the incident wave. For the two coupled modes, one can write the following Hamiltonian: 
\begin{figure}[t!]
\includegraphics[width=0.95\columnwidth]{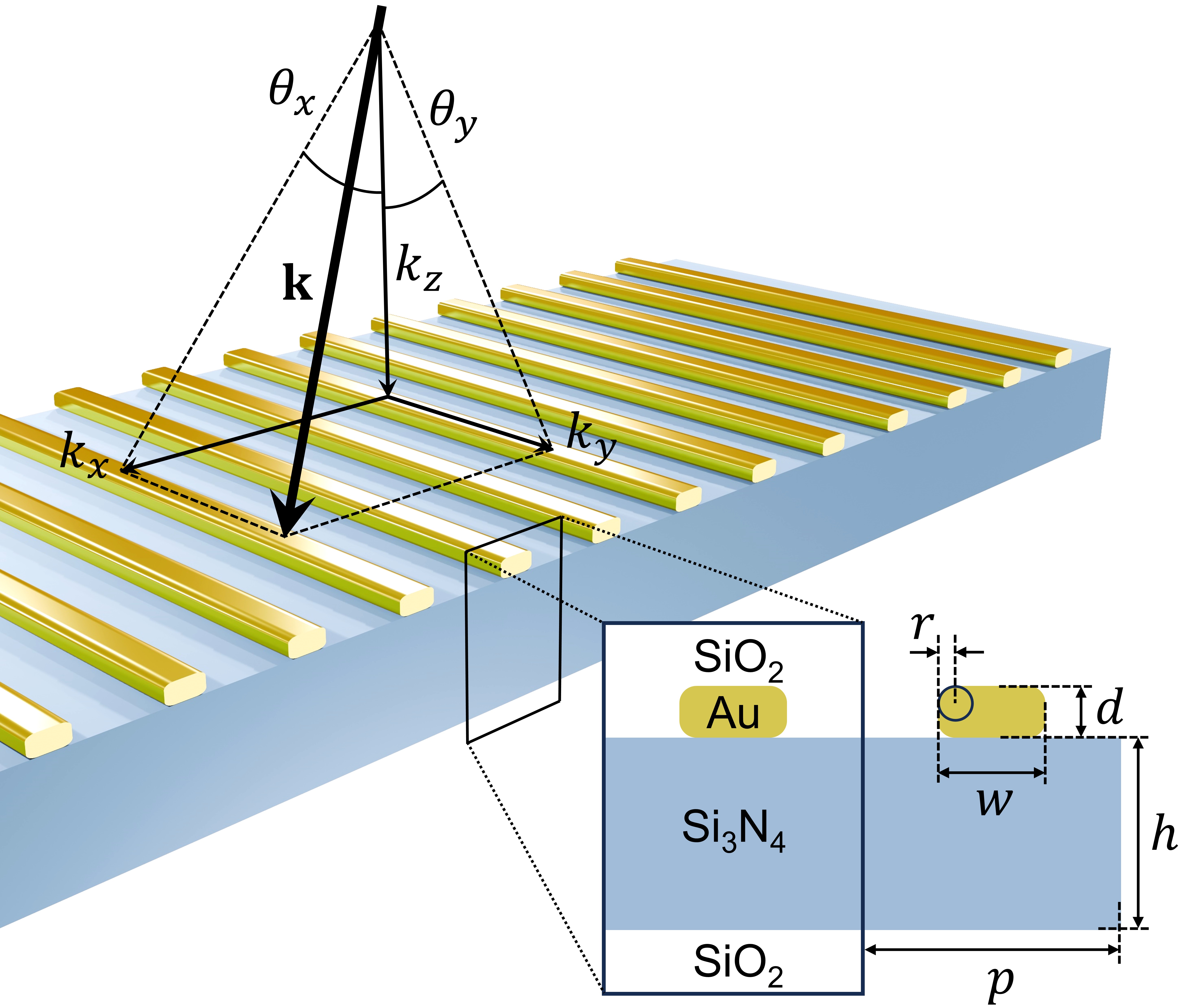}
	\caption{The structure of the proposed nonlocal non-Hermitian metasurface. The inset shows a cross section of the structure and defines its materials and dimensions $p$, $h$, $w$, $d$, and $r$. The incident wave vector $\mathbf{k}$ is defined by its angles $\theta_x$ and $\theta_y$ with respect to the normal to the metasurface plane. The vector component of $\mathbf k$ along the normal is $k_z$, and $k_x$ and $k_y$ are perpendicular and parallel to the gold bars of the grating, respectively.}\label{Fig2}
\end{figure}
\begin{align}
	H = \begin{pmatrix}
		\omega_0 &v_gk \\
	 v_gk & \omega_0+i\gamma
	\end{pmatrix}.
\end{align}
\rrrv{This Hamiltonian (with $i$ replaced with $-i$, depending on the convention) is rigorously derived in Ref.~\cite{zhen2015spawning} and its Supplementary Information, where it is used to describe the coupling between two modes of a photonic crystal slab. The Hamiltonian assumes one lossy and one lossless mode that are degenerate and coupled to each other at the $\Gamma$-point. Similar Hamiltonians have been used to describe the optical modes in coupled waveguides emulating the non-Hermitian physics~\cite{guo2009observation,weimann2017topologically,longhi2018quantum}.  In Ref.~\cite{rodriguez2016classical}, it has been demonstrated that, under certain conditions, such Hamiltonians can also be used to describe systems of coupled harmonic oscillators (see Eq. (10) in Ref.~\cite{rodriguez2016classical}). This makes the Hamiltonians applicable to various resonant systems, including photonic crystals and metasurfaces~\cite{kolkowski2023enabling}. In such cases, the lossy and lossless modes may correspond to bright and dark GMRs with radiative loss (or its absence) dependent on the coupling to the radiation continuum, which is governed by the symmetry selection rules~\cite{overvig2020selection}.}

\rrrv{The coupling between the two modes gives rise to two eigenmodes of the system with eigenfrequencies given by (see Eq. (2) in Ref. \cite{zhen2015spawning})}
\begin{align}
	\omega_\pm = \omega_0 +i\frac{\gamma}{2}\pm v_g\sqrt{k^2-k_c^2},
\end{align}
where $k_c=\gamma/2v_g$. The real and imaginary parts of the eigenfrequencies are plotted in Fig. 1(a). For $|k| < k_c$, the two modes are degenerate. At $|k| = k_c$, we have an exceptional point, at which the modes split in frequency. The splitting becomes more significant at increasing $|k| > k_c$. Since at $|k| \geq k_c$, the modes are strongly coupled to each other, their attenuation rates (imaginary parts of $\omega$) are equal~\cite{rodriguez2016classical}. The quality factors ($Q$) of the modes are defined as $Q = \frac{1}{2}\operatorname{Re}(\omega)/\operatorname{Im}(\omega)$. At $k$ = 0 (i.e., at the $\Gamma$ point), the $Q$ factor of the dark mode diverges to infinity. The mode is therefore a totally isolated BIC.

%-----------------------------------------------------table 1
\renewcommand{\arraystretch}{1.3} 
\begin{table}[b]
\caption{\label{table1} Geometrical parameters of the metasurface for which the band structure shown in Fig. \ref{Fig1}(b) was calculated.}
\begin{ruledtabular}
\begin{tabular}{cccccc}
Parameters&$p$&
$h$ &
$w$ &
$d$ &$r$\\
\colrule
Value (nm)&330 & 340 & 110 & 55 & 20\\ 
\end{tabular}
\end{ruledtabular}
\end{table}

If $\gamma$ results mainly from the radiation loss, then the imaginary part of the eigenfrequency is directly linked to the efficiency of coupling between the mode and the incident wave. This fact can be used to realize spatial filtering in a realistic photonic structure, e.g., a metasurface based on a metal grating coupled to a waveguide (see Fig. \ref{Fig2}). In such a metasurface, the guided modes of the waveguide are perturbed by the grating, which folds their dispersion curves into the light cone through diffraction~\cite{kolkowski2023nonlinear}. The resulting GMRs are exposed to radiation loss  which, however, can be suppressed by their destructive interference in the far-field, giving rise to a symmetry-protected BIC at the $\Gamma$-point~\cite{overvig2020selection}. This makes such GMRs a perfect platform for implementing the non-Hermitian model discussed above.  Simultaneously, the spatial overlap between the grating bars and the near-fields of the interfering modes affects the absorption loss of the GMRs~\cite{kolkowski2023enabling}, adding to the tunability of spatial filters that we want to design.

\section{Results and Discussion}

\subsection{Design of a metasurface for one-dimensional spatial filtering}
In order to demonstrate the non-Hermitian character and the capabilities of the proposed GMRs for spatial filtering, we perform frequency-domain electromagnetic simulations using the commercial finite-element software COMSOL Multiphysics. The considered metasurfaces consist of gold bars placed on a Si$_3$N$_4$ slab waveguide embedded in SiO$_2$ (see the inset of Fig. \ref{Fig2}). We choose the bars to be oriented along the $y$-axis and repeated periodically along the $x$-axis. 
%We assume that the grating and the waveguide  are surrounded on both sides (top and bottom) by thick SiO$_2$ layers. We consider the metasurfaces operating in the visible spectral range, meaning that the geometric parameters are in the nanoscale regime. 
The optical constants of the materials used in the simulations are taken from \cite{polyanskiy2024refractiveindex}. Both SiO$_2$ and Si$_3$N$_4$ are considered to be non-absorbing in the studied spectral range.

%-----------------------------------------------------Fig 3
\begin{figure}[t]
	\includegraphics[width=0.9\columnwidth]{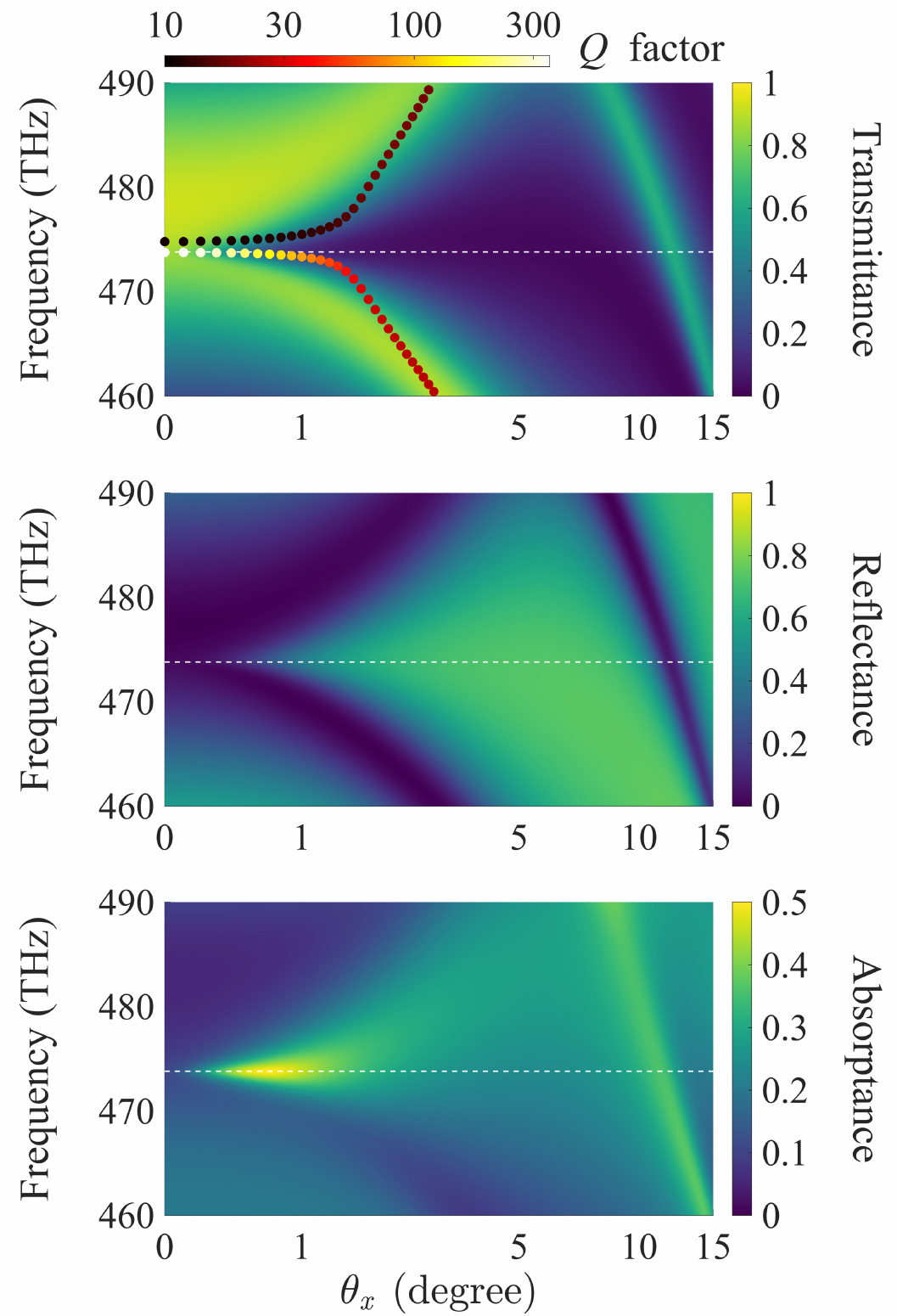}
	\rv{\caption{Numerically calculated transmittance, reflectance, and absorptance spectra of the TM metasurface with the geometric parameters specified in Table \ref{table2}. The eigenfrequencies (real part) of two GMRs of the structure are shown in the transmittance plot, and their $Q$ factors are encoded in the color. White dashed lines in each plot indicate the frequency for which the metasurface is designed to act as a spatial filter (the He-Ne laser frequency of 473.67 THz). }\label{Fig3}}
\end{figure}

Figure \ref{Fig1}(b) shows the numerically calculated band structure for a metasurface in which GMRs form a non-Hermitian degenerate flat band interacting with TM-polarized incident light (with the electric field lying in the $xz$-plane; we refer to such a metasurface as a ``TM metasurface''). The flat band is tuned to match the He-Ne laser frequency, $473.76$ THz ($\lambda=632.8$ nm). In the simulations, we assume the radius $r$ of the metal surface curvature at the corners of the bars (see Fig. \ref{Fig2}) to be fixed at 20 nm, which is typical for nanofabricated metal samples. \rrrv{Other geometric parameters selected for the structure to operate at $\lambda$ = 632.8 nm are listed in
Table~\ref{table1}. Remarkably, the band structure in Fig. \ref{Fig1}(b) is very similar to that obtained with our simple
analytical model and presented in Fig. \ref{Fig1}(a). The comparison shows that the curves fit each other
at $\omega_0$ = 2$\pi\times$473.76 THz and $k_c$ = 3.4$\times10^5$ m$^{-1}$ $\approx$ 0.03$\times$2$\pi/\lambda$. The parameters in both the analytical
model (Eqs. (1) and (2)) and the numerical model (metasurface design) were chosen manually to
achieve a good agreement between Figs. \ref{Fig1}(a) and \ref{Fig1}(b). However, one could automate the fitting
by defining appropriate figures of merit, e.g., to quantify the “band flatness”, and make it possible
to numerically optimize the structures, e.g., with the help of machine learning.} Note that, in Fig.~\ref{Fig1}(b), we have replaced $k$ by the in-plane momentum component $k_x$, which is relevant for the dispersion of GMRs confined to the metasurface plane. Moreover, $k_x$ is directly linked to the incidence angle $\theta_x$ of light coupling to GMRs (see Fig. \ref{Fig2}).
%The polarization of modes presented in Fig. \ref{Fig3} is transverse magnetic (TM), where the magnetic field is along $z$-axis. 
%The dispersion shows two degenerate GMRs forming a flat band near the $\Gamma$ point ($k_x$ around $0$). As $k_x$ increases, the two branches split at the exceptional point ($k_x\approx0.035\, \pi/p$) symmetrically with respect to the flat band. The imaginary part of the eigenfrequency in Fig. \ref{Fig3}(b) reveals the attenuation constants of the GMRs, showing that the loss of the low-$Q$ GMR decreases, while that of the high-$Q$ mode increases when we move along the $k_x$-axis from the $\Gamma$ point towards the exceptional point. Even though the two modes split in real part at $|k|>k_c$,  their imaginary parts get close and two modes are still strongly coupled to each other. 

 Figure \ref{Fig3} shows the transmittance, reflectance, and absorptance of the TM metasurface as functions of light frequency and incidence angle $\theta_x$. \rrv{These quantities were calculated numerically, using COMSOL Multiphysics, for monochromatic plane waves at different frequencies and incidence angles, using the standard electromagnetic port boundary conditions.} The TM metasurface is optimized for low-pass spatial filtering at 473.76 THz. (indicated by the white dashed line in each plot). We have slightly modified some of the geometric parameters (see Table \ref{table2}) to make the resonance frequencies of the two GMRs slightly different from each other at the $\Gamma$-point. This has been found to improve the spatial filtering performance of the structure compared to the case of perfectly degenerate GMRs. The eigenfrequencies of the corresponding GMRs are shown as dots in the transmittance plot, and their $Q$ factors are encoded in the color of the dots.

\begin{table}[b]
\caption{\label{table2} Geometrical parameters of the metasurface for which the results shown in Fig. \ref{Fig3} were obtained.}
\begin{ruledtabular}
\begin{tabular}{cccccc}
Parameters&$p$&
$h$ &
$w$&
$d$ &$r$\\
\colrule
Value (nm)&330 & 340 & 119.6 & 52 & 20\\ 
\end{tabular}
\end{ruledtabular}
\end{table}

%-----------------------------------------------------Fig 4
\begin{figure}[t]
	\includegraphics[width=0.88\columnwidth]{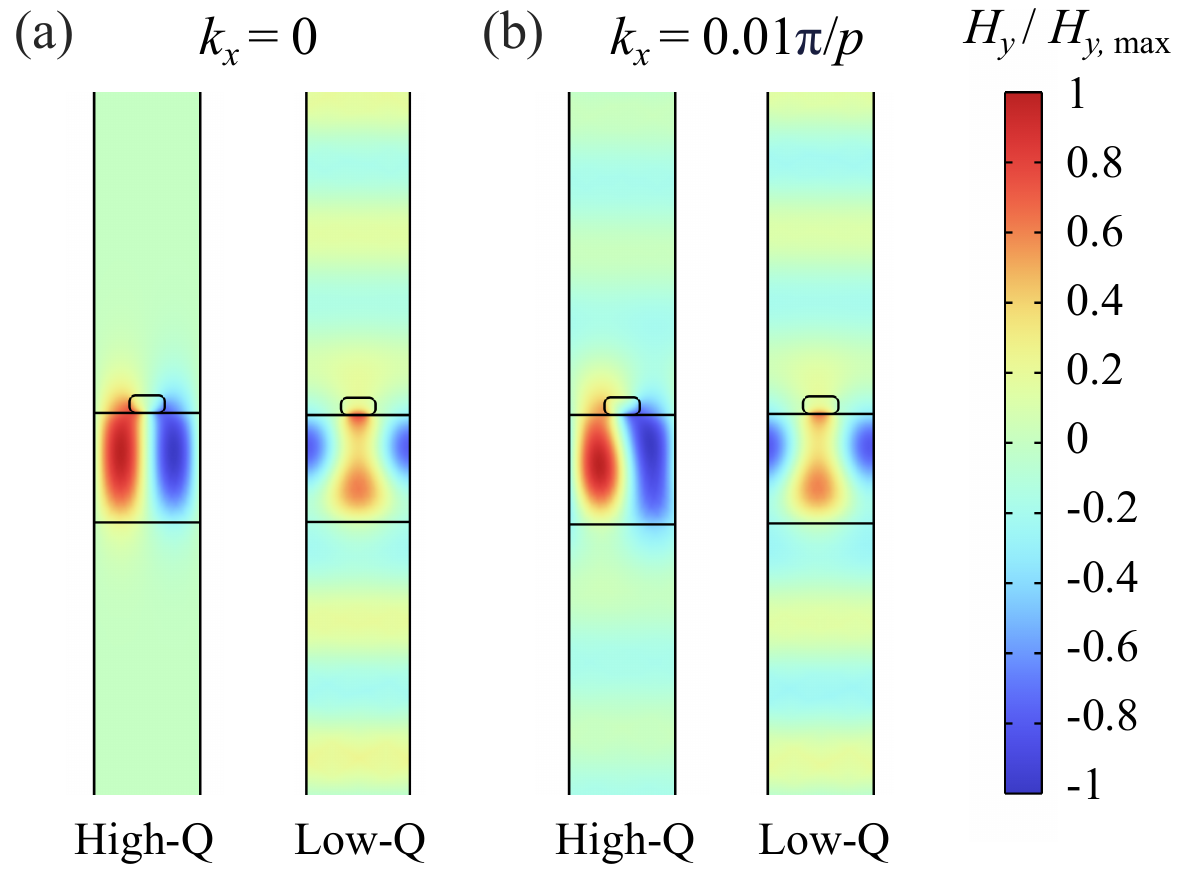}
	\vspace{-0.3cm}
	\caption{Instantaneous magnetic field profiles, $\operatorname{Re}(H_y)$, of the high-Q and low-Q GMRs at (a) $\Gamma$-point and (b) $k_x=0.01\pi/p$. The phase of the field was chosen to maximize the plotted amplitude.}
	\label{Fig4}
\end{figure}

%-----------------------------------------------------Fig 5
\begin{figure}[t]
	\includegraphics[width=0.97\columnwidth]{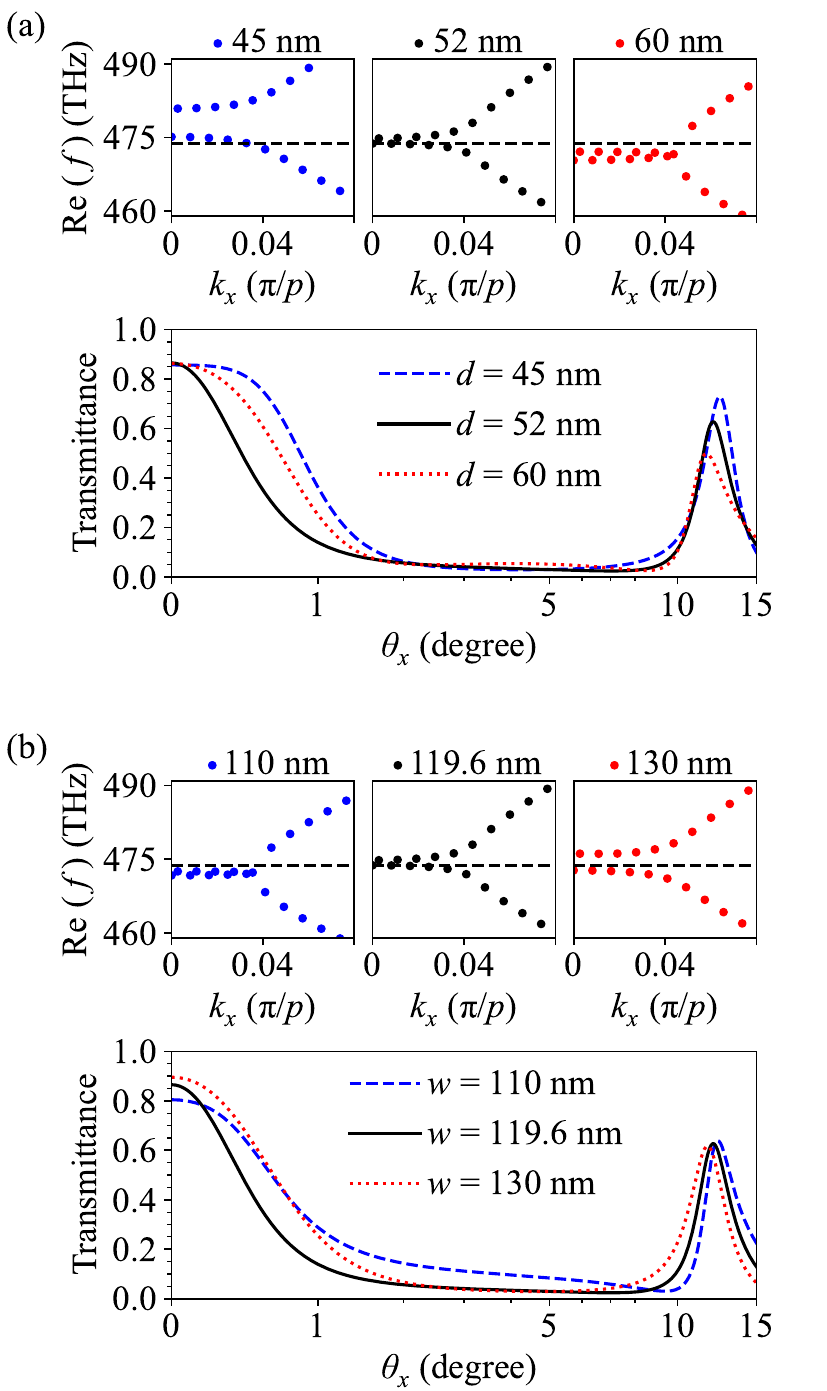}
	\caption{Non-Hermitian character of the band structure and spatial filtering performance of the TM metasurface depending on (a) thickness $d$ and (b) width $w$ of the gold bars. The transmittance curves are plotted for the same frequency as in the previous figures (473.67 THz). }\label{Fig5}
\end{figure}
 
The spatial filtering mechanism works as follows. If the incident light has a frequency that is far from the resonance frequencies of the two GMRs, it can only weakly be coupled to the structure. Under such off-resonance conditions, the structure is highly reflective for the polarization perpendicular to the grating bars, i.e., the TM polarization. Near the normal incidence (small $\theta_x$), the GMRs form a nearly degenerate flat band, similar to those in Fig. 1(a) at $0<|k|/k_c<1$ (with $k$ in the analytical model corresponding to $k_x$ in the case of the metasurface). At normal incidence ($k$ = 0, $\theta_x$ = 0), the narrowband dark GMR (corresponding to the mode shown by the blue line in Fig. 1(a)) is a BIC that cannot be excited due to a symmetry mismatch with the incident wave. In contrast, the bright GMR (corresponding to the mode shown by the red line in Fig. 1(a)) can be excited, transmitting the light to the other side of the metasurface. This is similar to the extraordinary optical transmission through nanoaperture arrays mediated by surface plasmon polariton Bloch modes~\cite{ebbesen1998extraordinary}. Note that the bright mode has a wider spectrum that covers also the design frequency of 473.76 THz. At a small $\theta_x$, the intensity in the bright GMR stays low, which is typical for low-$Q$ resonators, and therefore, the absorption loss remains insignificant. When $\theta_x$ increases, the dark GMR becomes a high-$Q$ quasi-BIC and the intensity in it grows fast, like in an externally driven high-$Q$ resonator. This leads to a significant increase of absorption. At the same time, the bright GMR becomes less lossy, so that its interaction with the incoming light becomes stronger, increasing the overall absorption even more. Hence, using a metasurface of this type, the transmittance can be made to change sharply from a high to a low value upon increasing $\theta_x$, which can be used to create a narrowband low-pass spatial filter. Using a non-Hermitian flat band ensures that the modes controlling the transmittance overlap with the design frequency over a large range of $\theta_x$. Beyond the flat band, the structure is off-resonant and highly reflective, which means that the transmittance stays low and the low-pass filtering function is maintained.

Figure \ref{Fig4} provides further insight into the spatial filtering mechanism, showing the instantaneous distribution of the magnetic field ($H_y$) of the two GMRs at $k_x$ = 0 and $k_x = 0.01\pi/p$. For the high-$Q$ GMR at the $\Gamma$-point, the field distribution is antisymmetric, which makes it perfectly decoupled from the free-space radiation. This is in contrast to the low-$Q$ GMR, which exhibits a symmetric field profile, making it prone to radiation loss. When the designed metasurface is illuminated by a normally incident plane wave matching the flat-band frequency, only the low-$Q$ GMR is excited. On the other hand, the excitation efficiency is in this case low and the absorption loss caused by the excitation is negligible. Tuning the structure parameters such that the GMRs have slightly different frequencies at the $\Gamma$-point allows us to efficiently control the absorption with the high-$Q$ GMR. As $k_x$ increases, the high-$Q$ GMR is no longer isolated, allowing more light to be coupled to the metasurface and more optical power to be dissipated. Combined with the increasing power of the low-$Q$ GMR, the overall absorption increases, enhancing the sharpness of the spatial filtering. 

Although in Table \ref{table2} we have specified one of the parameters with a high precision ($w$ = 119.6 nm), the proposed metasurfaces are very robust against deviations from the optimized geometry. Figure \ref{Fig5} shows that the metasurface retains most of its non-Hermitian character and its filtering functionality despite changing parameters $d$ and $w$ by as much as 10 nm off the optimized values. This demonstrates a quite acceptable fabrication tolerance of our design.  \rrv{Moreover, considering possible high-power illumination, the variation in the refractive index of $\rm Si_3N_4$ induced by the optical Kerr effect can be on the order of $10^{-4}$ \cite{soltani2019laser, moss2013new}. However, in such a case, the resulting shift in the flat band frequency would be much smaller than that caused by the considered geometrical deviations. On the other hand, excitation of high-Q resonances could lead to accumulation of the nonlinear effects, allowing them to modify the optical response in a noticeable way. We believe that the proposed metasurfaces could be redesigned to maximize such effects and to use them for nonlinear information processing.}

For TE-polarized incident light (with the electric field vector lying in the $yz$-plane), the structure can be designed to  act as a low-pass spatial filter operating in the reflection mode. At normal incidence, it should play the role of a highly reflective wire-grid polarizer and the absorption by the structure would be low due to the same reasons as in the example above. When the incidence angle increases, the absorption would increase as well, because light would be coupled to the GMRs that both become lossy. At even larger incidence angles, the system would cross the exceptional point and light would become off-resonant, resulting in low reflectance and high transmittance. Note that such a reflective low-pass filter can be used 
as a transmissive high-pass filter for TE-polarized light, which we demonstrate at the end of this paper. For the two operation regimes, the structures should be optimized independently.

%-----------------------------------------------------Fig 6
\begin{figure}[t]
	\includegraphics[width=0.95\columnwidth]{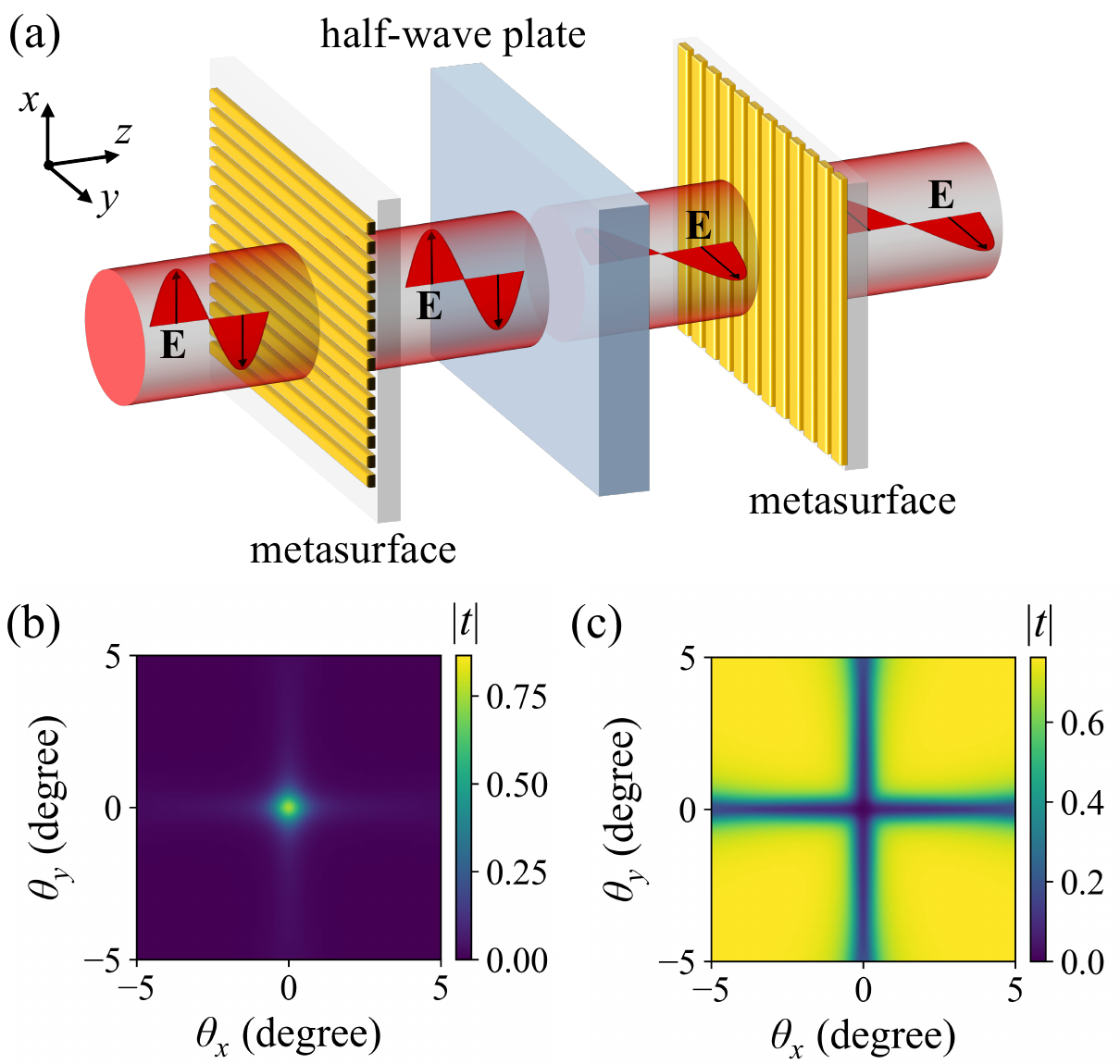}
	\caption{(a) \rv{Schematic of a 2D  low-pass spatial filter} and (b) its \rv{angular transmission spectrum}. The device is designed for linearly polarized incident light, TM-polarized with respect to the first metasurface. Both metasurfaces in the system operate using the TM-polarized GMRs. The same configuration can be used for 2D high-pass spatial filtering by replacing the TM-optimized metasurfaces with those optimized for TE polarization and employing TE-polarized GMRs. The \rv{angular transmission spectrum} of such a system is presented in (c). }\label{Fig6}
\end{figure}

%To implement the concept of utilizing properties near the exceptional point for achieving sharp spatial filtering, we design a one-dimensional nonlocal, non-Hermitian metasurface as shown in Fig. \ref{Fig2}, in which the inset illustrates the geometry and parameters for modeling the metasurface.  

\subsection{Two-dimensional spatial filtering}

Let us next demonstrate the possibility to create a transmissive low-pass spatial filter operating in the two-dimensional momentum space. We propose to combine two orthogonally-oriented TM metasurfaces with a half-wave plate positioned between the metasurfaces. The device is illustrated in Fig. \ref{Fig6}(a). The incident light must be linearly polarized, and the polarization must be perpendicular to the grating bars of the first metasurface (TM polarization). The first metasurface filters the incident light along the $x$-direction, and the second one along the $y$-direction. The half-wave plate rotates the polarization plane by 90 degrees. The \rv{angular transmission spectrum} of the setup is shown in Fig. \ref{Fig6}(b). It was calculated numerically by varying both angles $\theta_x$ and $\theta_y$ (see Fig. \ref{Fig2}) of the incident plane wave. \HL{The polarization for each incident wave is defined as the projection of \rv{the $x$-polarized electric field onto the transverse plane of}  the spatial Fourier component of the input beam\rv{, corresponding to the paraxial approximation. As an example, we} consider an $x$-polarized Gaussian beam propagating along \rv{the} $z$-axis, \rv{with the electric field vector given by $\mathbf E = \hat{\mathbf x} E_0(x,y)$  at the beam waist.} To analyze the beam in the spatial frequency domain, we expand the Gaussian field profile $E_0(x,y)$ into its plane wave components $E(\mathbf k)$ \rv{and assign the projected polarization to each of them, i.e., $\mathbf E(\mathbf k) = \hat{\mathbf e}_t E_t(\mathbf k)$, where $ \hat{\mathbf e}_t$ is the projection of $\mathbf x$ onto the transverse plane ($\hat{\mathbf e}_t \perp \mathbf k$).}} The spectrum is seen to be nearly circularly symmetric\rv{, being peaked} at normal incidence. The angular bandwidth of the filter is on the order of 1 degree, as expected.

\HL{Similarly, for a 2D high-pass filter based on TE metasurfaces, a $y$-polarized Gaussian beam is treated using the same method as described above. The metasurface designed for this filter is described later on in the paper. The angular transmission spectrum of the system is shown in Fig. \ref{Fig6}(c). It rejects the plane-wave components with either $\theta_x$ or $\theta_y$ close to 0. When both incidence angles are} larger than 1 degree, the system is highly transmissive.

\begin{figure}[t!]
	\includegraphics[width=1\columnwidth]{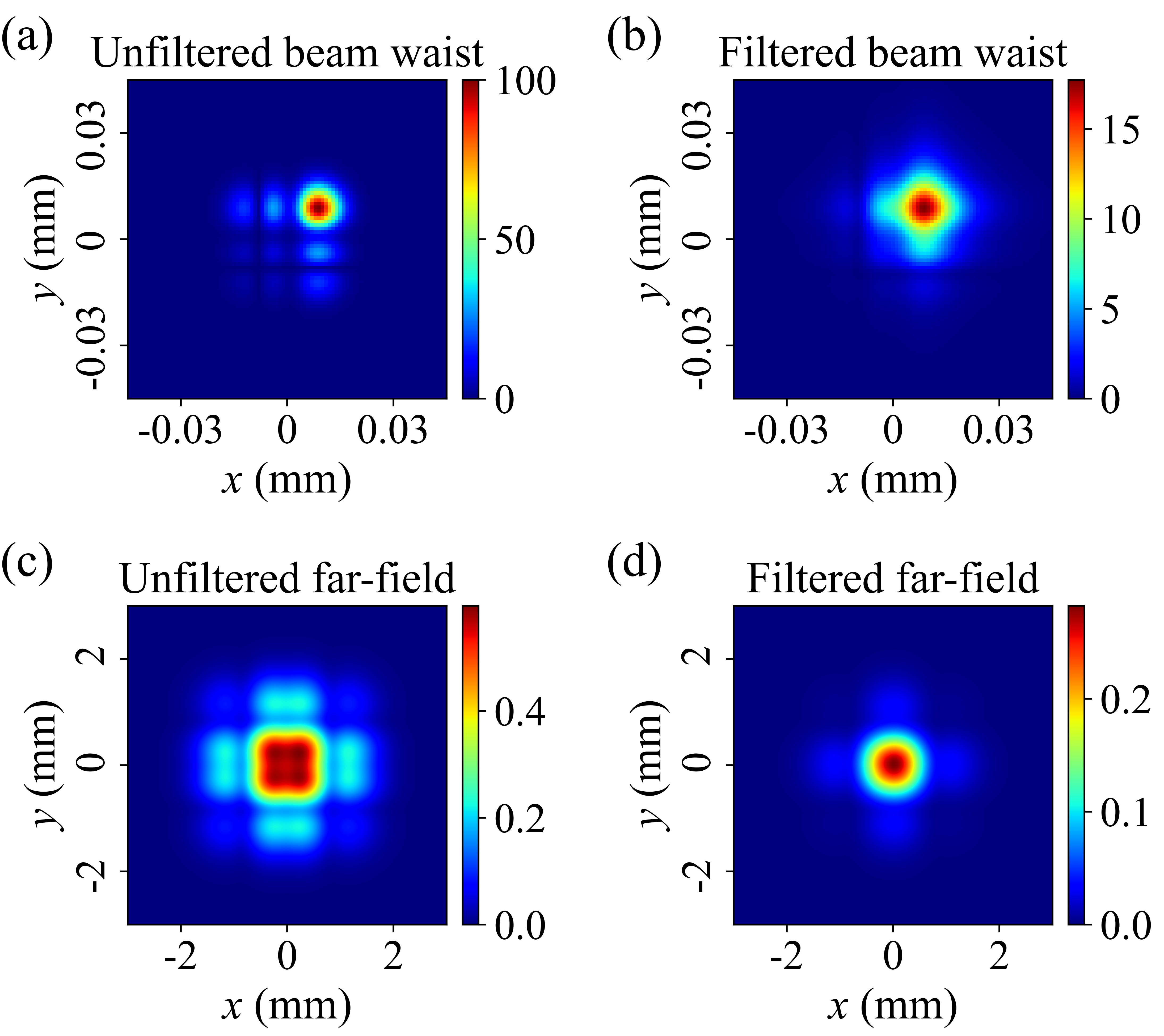}
	\rv{\caption{The transverse field profiles  of a multimode Hermite-Gaussian beam at the beam waist (a) before and (b) after filtering. In the far-field (at a distance of 30 mm from the beam waist), the beam profiles are shown in (c) and (d), respectively. }\label{Fig7}}
\end{figure}

\begin{figure}[t]	\includegraphics[width=0.9\columnwidth]{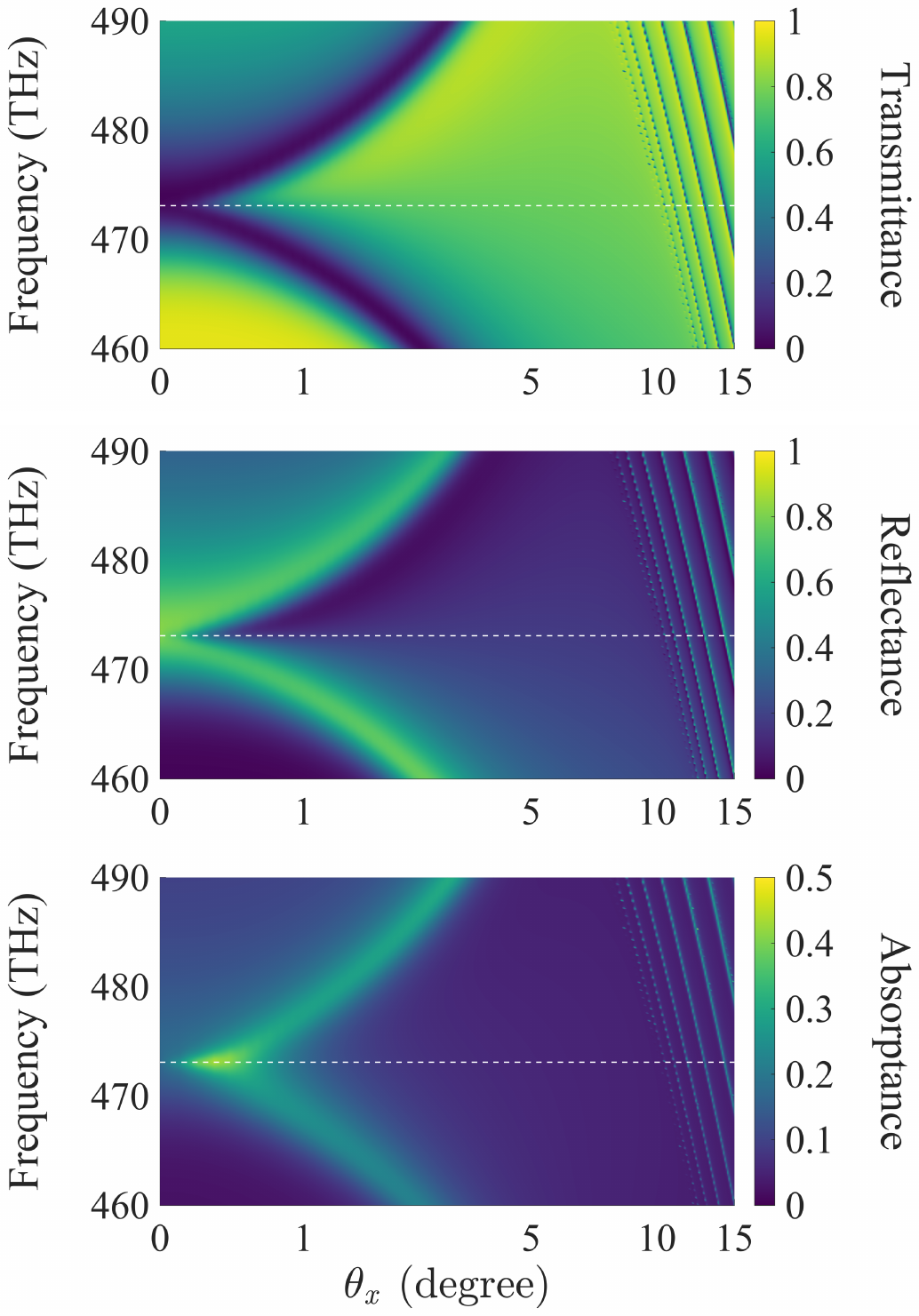}
	\rv{\caption{Transmittance, reflectance, and absorptance spectra of the metasurface acting as a transmissive high-pass spatial filter for TE-polarized incident light. The parameters of the metasurface are given in the text.}\label{Fig8}}
\end{figure}

\begin{figure*}[t!]
	\includegraphics[width=0.75\textwidth]{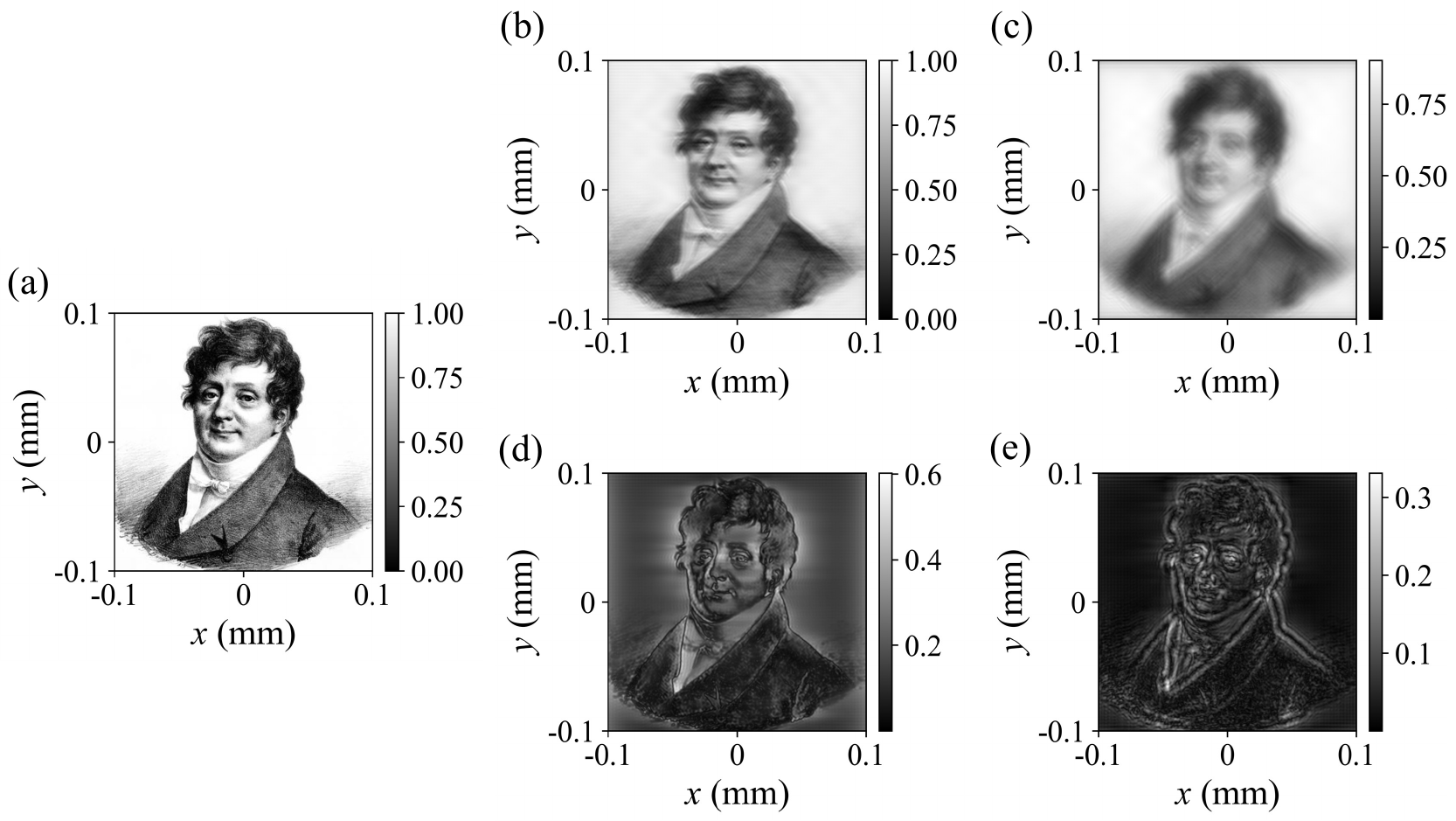}
	\vspace{-0.3cm}
	\caption{\HL{Demonstration of spatial filtering }by the nonlocal non-Hermitian metasurfaces designed in this work. (a)  Portrait of Joseph Fourier used as the input image. The output images are shown in (b-e), resulting from filtering by (b) 1D low-pass filter based on a single TM metasurface, (c) 2D low-pass filter based on the setup in Fig. \ref{Fig6}(a), employing two TM metasurfaces (the corresponding angular \HL{transmission} spectrum is shown in Fig.~\ref{Fig6}(b)), (d) 1D high-pass filter based on a single TE metasurface, and (e) 2D high-pass filter similar to the one in Fig. \ref{Fig6}(a), but with two TE metasurfaces and incident TE-polarized light (the corresponding angular \HL{transmission}  spectrum is shown in Fig.~\ref{Fig6}(c)). }\label{Fig9}
\end{figure*}

We adopt a Fourier-optics-based numerical approach to quantitatively evaluate the filtering performance of the proposed devices. The filter shown in Fig. 6(a) is tested using a multimode Hermite-Gaussian beam (a sum of 16 modes with equal powers, $\sum_{l,m=(0,0)}^{(3,3)}U_{l, m}$) as the input. \HL{Each plane wave component of the input beam propagates through the designed 2D filtering device, experiencing a unique complex transmission coefficient, $t$ (see the plot of $|t|$ in Fig. 6(b)). The transmitted field is then reconstructed by superposing the altered  components. The far-field divergence angle and beam waist radius of the fundamental mode are set to 1.5$^\circ$ and 7.7 $\upmu$m, respectively.}   We tune the geometrical parameters of the metasurfaces to $w=130$ nm and $d=52$ nm, such that the filtering profile closely resembles a Gaussian distribution. The transverse field profiles before and after the 2D low-pass filtering are presented in Fig. \ref{Fig7}. Based on the corresponding intensity profiles, we calculate the beam $M^2$-factor, far-field divergence angle $\Delta\theta$, and Rayleigh range $z_\mathrm R$ \cite{hooker2010laser}, as well as the power transmittance of the fundamental mode, $T_{00}$, defined as

\begin{align}
	T_{00} = \frac{P_{00,\rm out}}{P_{00,\rm in}} = \frac{|\iint_{-\infty}^\infty U_{00}^*(x,y)U_{\rm out}(x,y)\mathrm dx\mathrm dy |^2}{|\iint_{-\infty}^\infty U_{00}^*(x,y)U_{\rm in}(x,y)\mathrm dx\mathrm dy |^2},
\end{align} 
\rrrv{where $P_{00,\rm in/out}$ is the power fraction of the fundamental mode in the total field profile $U_{\rm in/out}(x, y)$. The function $U_{00}(x, y)$ in the overlap integrals of the above equation is the normalized field profile of the fundamental mode ($l=0,m=0$). For example, if $U_{\rm in/out}(x, y)$ is a superposition of modes $U_{00}$ and $U_{01}$ with amplitudes $A$ and $B$, respectively (i.e., $U_{\rm in/out}(x,y) = AU_{00}(x,y) + BU_{01}(x,y)$), the squared absolute value of the overlap integral is equal to $|A|^2$, which follows from the orthonormality of functions $U_{00}(x,y)$ and $U_{01}(x,y)$.} The beam parameters \rv{before and after} filtering are listed in Table \ref{table3}. \rv{The filter significantly improves the beam quality, as the $M^2$ value is reduced from \HL{$1.84$ to $1.23$}, while preserving a half of the fundamental mode power.} A high-quality performance of the device is confirmed by the far-field distribution in Fig. \ref{Fig7}(d), exhibiting a nearly Gaussian profile with essentially removed transversely diverging ($k_t>0.01\pi/p$)  higher-order modes.

\begin{table}[t]
	\caption{\label{table3} Parameters of the unfiltered and filtered  multimode Hermite-Gaussian beams ($\sum_{l,m=(0,0)}^{(3,3)}U_{l, m}$)}
	\begin{ruledtabular}
		\begin{tabular}{ccccc}
			Parameters&$M^2$&
			$\Delta\theta$ ($^\circ$)&
			$z_\mathrm R$ (mm)&
			$T_{00}$\\
			\colrule
			\HL{Unfiltered}&1.84 & 2.16 & 0.48 &1.00 \\ 
			\colrule
			\HL{Filtered}& 1.23 & 1.24 & 0.65 & 0.57\\ 
		\end{tabular}
	\end{ruledtabular}
\end{table}

Finally, we use the same Fourier-optics-based method to demonstrate the capability of the proposed metasurfaces for image processing, including edge detection \cite{silva2014performing, zhou2019optical, wan2020optical, komar2021edge, bi2023wideband, cotrufo2023dispersion, cotrufo2023polarization, tanriover2023metasurface}. For this purpose, in addition to the transmissive low-pass-filtering TM metasurface, we \rv{use} a transmissive high-pass-filtering metasurface operating on the TE-polarized light and using the TE-polarized GMRs. The optimal geometry parameters for this metasurface are $p$ = 370 nm, $h$ = 180 nm, $w$ = 210 nm, and $d$ = 40 nm.

Figure \ref{Fig8} presents the corresponding reflectance, transmittance, and absorptance spectra, demonstrating that\rv{,} at the selected frequency (473.67 THz), the metasurface exhibits a narrow reflective band near the $\Gamma$-point and becomes highly transmissive as the angle of incidence increases. This makes the metasurface an effective one-dimensional \rv{high-pass  filter}. Two-dimensional high-pass filtering can be achieved by assembling two such TE metasurfaces and a half-wave plate in the same way as the TM metasurfaces are assembled in Fig.~\ref{Fig6}(a). The resulting angular transmittance spectrum is shown in Fig.~\ref{Fig6}(c). As an example, we show in Fig.~\ref{Fig9} a portrait of Joseph Fourier filtered by our low-pass and high-pass spatial filters, both in 1D (using just one metasurface) and in 2D (using the whole device). The images become blurred after low-pass filtering, while the edges of the images become clearly visible after high-pass filtering.

\section{Summary}
To summarize, we have demonstrated the possibility to perform efficient spatial filtering by nonlocal  metasurfaces supporting a non-Hermitian flat band in the vicinity of a symmetry-protected bound state in the continuum. Such a peculiar band structure is achieved by coupling between the high- and low-$Q$ guided-mode resonances, which are tailored by the geometric parameters of the metasurface building blocks, i.e., the slab waveguide and the metallic grating. Due to the interaction of the incident light with the designed resonances, the proposed metasurfaces exhibit abrupt changes in the transmittance, reflectance, and absorptance on the way from normal to oblique incidence, allowing them to operate as either low-pass or high-pass spatial filters, depending on the light polarization and the polarization of the guided-mode resonances. Owing to their inherently nonlocal character, the optical properties of the proposed  metasurfaces do not depend on their lateral or longitudinal dispacement, making them nearly alignment-free compared to the conventional spatial filters. We also show that their filtering performance is relatively insensitive to the deviations of the geometrical parameters, which makes them highly feasible from the fabrication point of view. Their high performance demonstrated in this work, together with their practical feasibility and compactness, make them a promising candidate to become a widespread optical element, with applications ranging from \rv{optical interferometry} to coherent imaging. \rrv{Further extensions of the proposed nonlocal non-Hermitian metasurfaces, e.g., by introducing nonlinear and/or non-reciprocal phenomena, as well as involving the non-Hermitian skin effect \cite{yao2018edge, longhi2019probing,okuma2020topological}, could provide novel and exotic means for all-optical information processing.}

\vspace{0.5cm}
\section*{Acknowledgement}

The authors acknowledge support of the Research Council of Finland (Grants No. 347449 and 353758). We also acknowledge funding from the Research Council of Finland Flagship Programme, Photonics Research and Innovation (PREIN), decision number 346529. For computational resources, the authors acknowledge the Aalto University School of Science “Science-IT” project and CSC – IT Center for Science, Finland.

%\section*{Supplemental Information}
% The angle transforation from $(\theta, \varphi)$ to $(\theta_x, \theta_y)$ is by,
% \begin{align}
% 	\theta_x &= \sin^{-1}\left(\sin\theta\cos\varphi \right), \\
% 	\theta_y &= \varphi.
% \end{align}
%The angular transmittance spectrum of 2D high-pass filter based on TE metasurfaces is shown in Fig. \ref{Fig10}.

%\begin{figure}[htb!]
%	\includegraphics[width=0.6\columnwidth]{Fig10.pdf}
%	\caption{Angular transmittance spectrum of 2D high-pass filter.}\label{Fig10}
%	\vspace{-0.4cm}
%\end{figure}
\bibliography{apssamp}% Produces the bibliography via BibTeX.

\end{document}